\documentclass{PoS}

\title{Effects of extremely strong magnetic field on photon HBT interferometry}

\ShortTitle{Effects of extremely strong magnetic field on photon HBT interferometry}

\author{\speaker{Kazunori Itakura}\\
        Theory Center, IPNS, KEK, 1-1 Oho, Tsukuba, Ibaraki 305-0801, Japan\\
        E-mail: \email{kazunori.itakura@kek.jp}}

\author{Koichi Hattori\thanks{Current address: Institute of Physics and Applied Physics, Yonsei University, Seoul 120-749, Korea}\\
        Theory Center, IPNS, KEK, 1-1 Oho, Tsukuba, Ibaraki 305-0801, Japan\\
        E-mail: \email{khattori@post.kek.jp}}

\abstract{We discuss potential attainability by the Hanbury Brown and 
Twiss (HBT) interferometry of photons to probe the spacetime geometry of 
the primordial plasma
created by ultrarelativistic heavy-ion collisions. A possible effect to
distort the HBT image is due to an interaction between emitted photon
and an extremely strong magnetic field induced by the colliding heavy
nuclei in the peripheral collisions. We examine the effects of 
variation of the refraction index in the strong magnetic field that is
called the vacuum birefringence.}

\FullConference{The Seventh Workshop on Particle Correlations and Femtoscopy\\
		 September 20th to 24th, 2011\\
		 The University of Tokyo, Japan}

\newcommand{\bk}{{\bm{k}}}

\newcommand{\bq}{{\bm{q}}}
\newcommand{\bx}{{\bm{x}}}

\usepackage{bm}
\usepackage{graphicx}
\usepackage{amsmath,amssymb,mathrsfs}
\usepackage{cases}

\begin{document}
\section{Strong Magnetic Fields in Non-central HIC and Non-linear QED}

It has been recently recognized that extremely strong magnetic fields 
are created in non-central heavy-ion collisions at high energies. 
For example, the strength amounts to $|eB|\sim 10^4 {\rm MeV}^2 \sim m_\pi^2$ 
at RHIC (and even stronger at the LHC), which is $10^5$ times 
higher than the electron's critical magnetic field\footnote{Its electric 
analog is the Schwinger's critical electric field, beyond which $e^+e^-$ pair 
creation occurs.} 
$|eB_c| = m_e^2=0.25~{\rm MeV}^2$ with $m_\pi$ and $m_e$ being 
the pion and electron masses, respectively. 
This is easily understood in a simple estimate using 
the Lienard-Wiechelt potential for fast moving nuclei with 
large electric charges \cite{Kharzeev:2007jp}, 
but can be also reproduced numerically in 
simulations such as UrQMD \cite{Skokov:2009qp}. 
While the magnetic fields rapidly decrease 
(typically as a power of $t$) as two charges receed from each other, 
they still remain much stronger than the critical field during the lifetime of the quark-gluon plasma (QGP). 
Therefore, we need to consider the dynamics of QGP, 
such as creation and time evolution, 
under the influence of a strong magnetic field \cite{ItakuraPIFreview}. 

In such a strong magnetic field, we expect qualitatively new phenomena 
called ``nonlinear QED effects". Even if the coupling of the charge to 
the magnetic field is weak, strongness of the magnetic field compensates 
the weakness of the coupling and we need to sum up all orders with 
respect to the coupling. 
For example, every insertion of the external magnetic field to the electron 
propagator gives ${\cal O}(eB/m_e^2)$, and thus when $eB \gg eB_c=m_e^2$, 
higher order terms with arbitrary number of
insertions are equally important to 
be summed up. In such a situation, observables exhibit nonlinear dependence
on the magnetic field, which is thus called the nonlinear QED effects. 
Examples include the synchrotron radiation of gluons/photons emitted 
by quarks, real 
photon decay $\gamma \to e^+e^-$, photon splitting $\gamma\to \gamma + \gamma$,
and the vacuum birefringence of a 
photon, the last of which is the subject of this talk.
Since these nonlinear QED phenomena become prominent in the strongest magnetic
field, they can be used as tools of probing the early time, particularly 
pre-equilibrium, stages of the heavy-ion collision events.

\section{Vacuum birefringence of a photon}

Photons are emitted at all stages of heavy-ion collisions. 
Direct photons are created in the initial hard collisions, thermal 
photons are from QGP, and decay photons are from decays of (mainly) pions. 
In addition to these well-known photon sources, 
in the presence of strong magnetic 
fields, there will be synchrotron radiation from quarks. Since photons do not 
interact via strong interaction, they escape from the collision area, and 
carry the information of initial stages of heavy-ion collision events. 
However, when extracting such information, we have to be careful about the 
nonlinear QED effects on photons. In particular, the initial direct photons 
and thermal photons from the earliest time QGP will be affected 
by the magnetic field through the vacuum birefringence. 
This talk briefly describes 
theoretical framework of the vacuum birefringence and examines its influences 
on the photon HBT images in a simple model.

In an ordinary vacuum, 
the one-loop self-energy diagram of a photon propagator, 
or the polarization tensor $\Pi^{\mu\nu}(q)$, 
does not change the speed of light. However,  in the presence of a 
strong  magnetic field that selects a specific direction in space, 
the tensor $\Pi^{\mu\nu}(q,B)$ acquires additional terms that depend 
on $B$ (see Fig.~1). Consequently, a photon can have, in general, refractive 
indices different from unity just as if the vacuum behaves as a medium. 
The refractive indices of two physical propagating modes are not 
necessarily the same, thus this is called ``birefringence".

\begin{figure}
\begin{minipage}{0.5\hsize}
\begin{center}
\includegraphics[width=.8\textwidth]{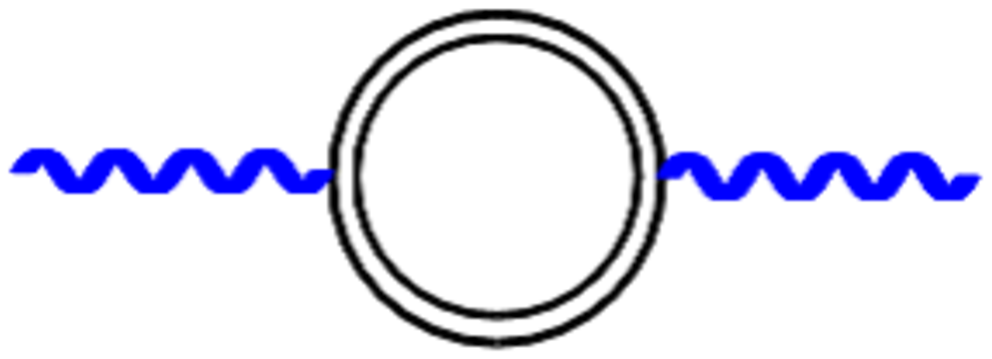}
\end{center}
\caption{Vacuum polarization in a magnetic field. Double line corresponds to 
a dressed electron propagator in the magnetic field.}
\label{fig1}
\end{minipage}
\quad\qquad 
\begin{minipage}{0.35\hsize}
\begin{center}
\includegraphics[width=.75\textwidth]{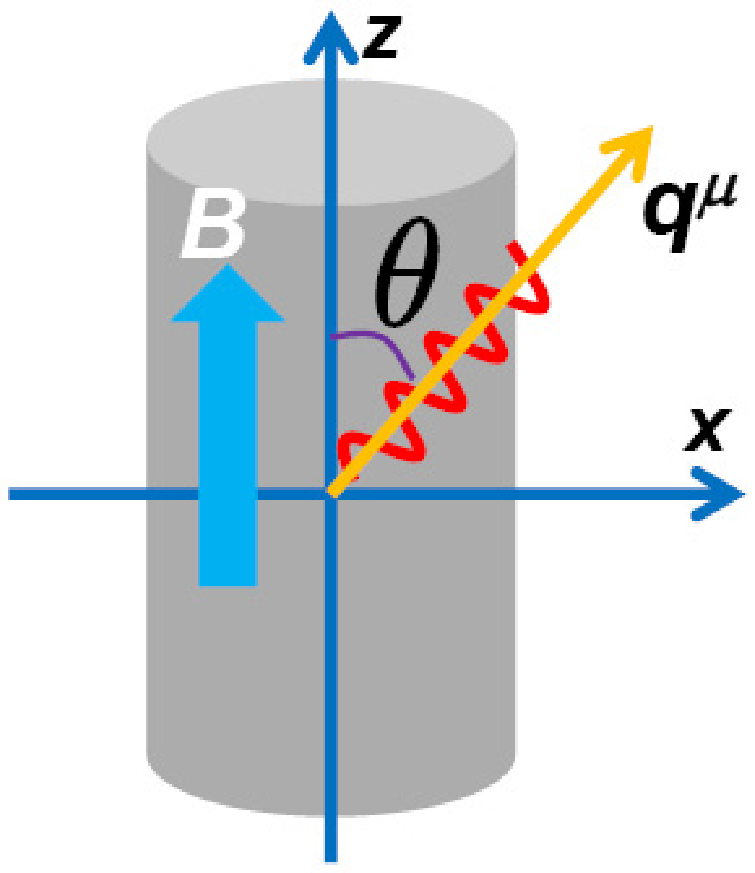}
\end{center}
\caption{A photon propagates in a strong magnetic field.}
\label{fig2}
\end{minipage}
\end{figure}

More specifically, the polarization tensor of a photon at one-loop level 
is given by 
\begin{equation}
i\Pi^{\mu\nu}(q,B)= (-ie)^2 (-1) \int \frac{d^4p}{(2\pi)^4} {\rm Tr} 
\left[
\gamma^\mu G(p,A) \gamma^\nu G(p+q,A)
\right]\, ,
\end{equation}
where $G(p,A)$ is the electron propagator in a magnetic field,
$iG^{-1}(p,A)=p\!\!\!\!/-eA\!\!\!\!\!/-m_e$. Note that this propagator
incorporates all-order contributions 
with respect to the external magnetic 
field. We assume that the magnetic field is oriented to the third direction
in spatial coordinates (see Fig.~2). 
With the special direction specified by the magnetic field, the Lorentz
structure of the polarization tensor $\Pi^{\mu\nu}$ is now modified as
\begin{equation}
\Pi^{\mu\nu}(q,B)= - \chi_0 \Big[q^2 \eta^{\mu\nu}-q^\mu q^\nu\Big]
- \chi_1 \Big[q^2_\parallel \eta_\parallel^{\mu\nu}-q_\parallel^\mu q_\parallel^\nu\Big]
- \chi_2 \Big[q^2_\perp \eta_\perp^{\mu\nu}-q_\perp^\mu q_\perp^\nu\Big]\, ,
\end{equation}
where the metric $\eta^{\mu\nu}=diag (1,-1,-1,-1)$ and the photon momentum 
$q^\mu=(q^0,q_\perp,0,q^3)$ are decomposed into two parts (parallel and 
perpendicular to the magnetic field); 
$\eta_\parallel^{\mu\nu}=diag(1,0,0,-1)$, 
$\eta_\perp^{\mu\nu}=diag(0,-1,-1,0)$, and  
$q^\mu_\parallel = (q^0,0,0,q^3)$, $q^\mu_\perp=(0,q_\perp,0,0)$, and 
$\chi_i$ ($i=0,1,2$) are scalar functions of $q_\parallel^2$, 
$q_\perp^2$ and $B$.
Without the magnetic field, we see that both $\chi_1$ and $\chi_2$ vanish, 
and that the last two terms are absent in a vacuum. 
With this polarization tensor, 
the Maxwell equation gets modified as 
$\big[ q^2 \eta^{\mu\nu}-q^\mu q^\nu - \Pi^{\mu\nu}(q,B)\big]A_\nu(q)=0$. 
By solving this equation, one obtains dispersion relations for two 
physical modes. 
One finds that, when $\chi_1$ and/or $\chi_2$ are nonzero, 
 these two are distinct from each other differently from those in a vacuum, 
$\omega^2={\bq}^2$. 
Defining the refractive index $n$ by 
\begin{equation}
n^2=\frac{|{\bq}|^2}{\omega^2},
\end{equation}
one obtains two different refractive indices 
\begin{equation}
n_1^2=\frac{1+\chi_0+\chi_1}{1+\chi_0+\chi_1\cos^2\theta},\qquad 
n_2^2=\frac{1+\chi_0}{1+\chi_0+\chi_2\sin^2\theta}\, ,
\end{equation}
where $q^\mu=(\omega,q_\perp,0,q_\parallel)
=(\omega, |{\bq}|\sin \theta,0,|{\bq}|\cos \theta)$ (see Fig.~2).  
Of course, when $\chi_1$ and $\chi_2$ are 
vanishing, both $n_1$ and $n_2$ reduce to unity. It should be noticed that 
the refractive indices $n_i$ depend on the angle $\theta$, and when $\theta=0$ 
(a photon propagates in parallel to the magnetic field),
both $n_1$ and $n_2$ are 1.

\section{Photon HBT interferometry in the presence of magnetic fields}

\begin{figure}
\begin{minipage}{0.45\hsize}
\begin{center}
\includegraphics[width=.7\textwidth]{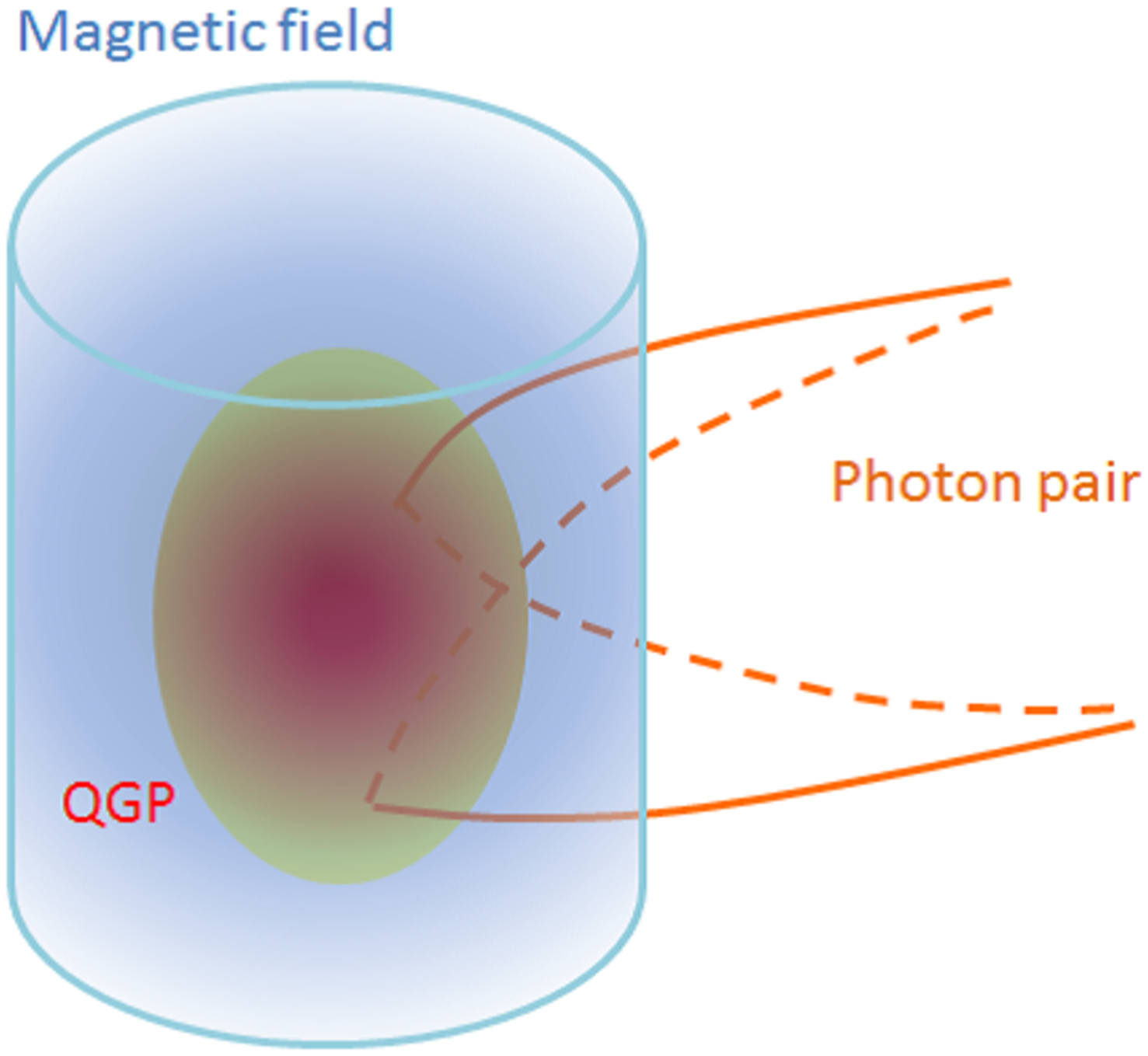}
\end{center}
\caption{
Photon HBT image will be distorted due to a modification of the optical path 
in the strong magnetic field. A pair of solid and dashed lines show an interference between two possible trajectories of a photon pair having identical polarizations. 
}
\label{fig3}
\end{minipage}
\qquad
\begin{minipage}{0.45\hsize}
\begin{center}
\includegraphics[width=.6\textwidth]{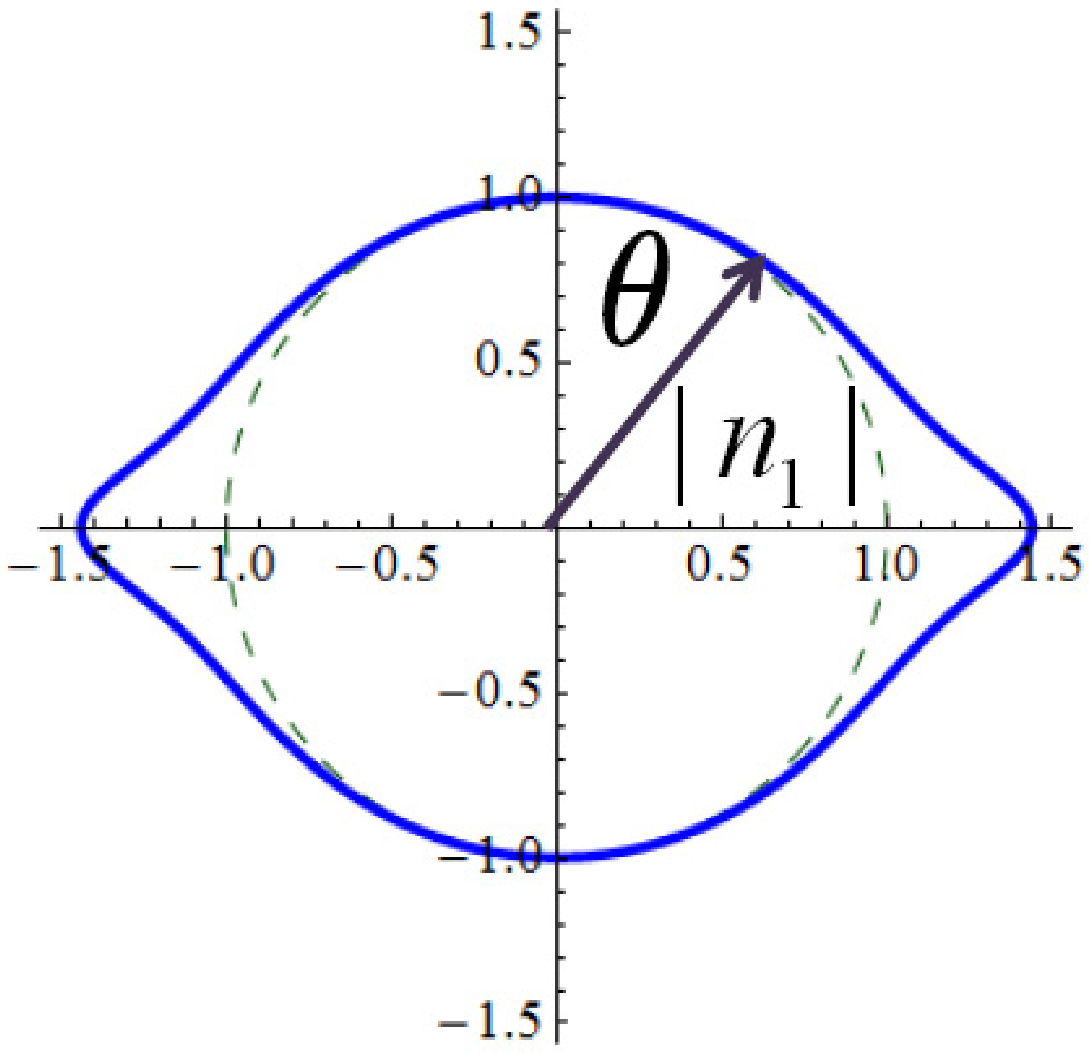}
\end{center}
\caption{The refractive index $n_1$ in 
the lowest-Landau-level approximation. 
}
\label{fig4}
\end{minipage}
\end{figure}

In the future experiments, it is quite important to measure the HBT 
interferometry of photons because the observables will contain
the information of the matter just after the heavy-ion collisions, 
hopefully, before the QGP is formed. This 
would be straightforwardly 
true when the magnetic field is absent. 
Photon HBT could directly give the information of the matter 
 at the moment of photon emissions. 
However, in the presence of a strong magnetic field which induces 
variation of photon refractive index, {\it photon HBT gives a distorted image}.
In order to correctly extract the early-time source distribution,
we need to understand how the image is distorted.

In the photon HBT interferometry, 
which has been originally developed in radio astronomy \cite{HBT}, 
the static source distribution can be 
related to the correlation function of  intensity $E_{\bk\lambda}$ of 
a photon with momentum $\bk$ and polarization $\lambda$ \cite{Gla}.
When no magnetic field is imposed in the source region, the relationship 
is quite straightforward: 
\begin{eqnarray}
C(\bk,\bq; \lambda_1,\lambda_2)&\equiv&
\frac{
\left\langle E^{(-)}_{\bk_1 \lambda_1} E^{(-)}_{\bk_2 \lambda_2}
        E^{(+)}_{\bk_1 \lambda_1} E^{(+)}_{\bk_2 \lambda_2}
\right\rangle }{
\left\langle E^{(-)}_{\bk_1 \lambda_1} E^{(+)}_{\bk_1 \lambda_1}
\right\rangle 
\left\langle E^{(-)}_{\bk_2 \lambda_2}E^{(+)}_{\bk_2 \lambda_2}
\right\rangle }\nonumber \\
&=& 1+\delta_{\lambda_1\lambda_2}\frac1N 
\left \vert \int d\bx\, S(\bx,\bk)\, e^{-i\bq\cdot \bx}\right\vert^2 \, ,
\end{eqnarray}
where $\bk=\bk_1+\bk_2$, $\bq=\bk_1-\bk_2$ and $S(\bx,\bk)$ is the static 
source distribution, and $N$ is a normalization factor.
Namely, the correlation function is just given by the Fourier transformation 
of the static source distribution. 
However, in the presence of a strong magnetic field, photon propagation 
is modified, and its trajectory will be bent due to the spatial profile 
of the magnetic field. Consequently, 
led by a phase distortion, the relationship between the static 
source and the correlation function becomes nontrivial. This situation 
is very similar to distortion of the HBT image of pions in the presence 
of a mean-field cloud \cite{HM}. One can formulate in a similar way as 
Ref.~\cite{HM}, and the relationship between $C(\bk,\bq;\lambda_1,\lambda_2)$ 
and $S(\bx,\bk)$ reads 
\begin{equation}
C=1+\delta_{\lambda_1\lambda_2}\frac1N \left\vert 
\int d\bx\, S(\bx,\bk)\, e^{-i\bq\cdot \bx +i\delta\psi(\bx; \bk, \bq)
-\Gamma(\bx;\bk,\bq)}
\right\vert^2 \, ,
\end{equation}
where $\delta\psi$ and $\Gamma$ are, respectively, 
difference of the eikonal angles of two photons 
$\delta\psi(\bx;\bk,\bq)=
\psi(\bx,\bk_1)-\psi(\bx,\bk_2)$ and sum of the extinction coefficients 
$\Gamma(\bx; \bk,\bq)=\gamma(\bx,\bk_1)+\gamma(\bx,\bk_2)$. 
Note that the eikonal angles and extinction coefficients are obtained 
as integrals of the real and imaginary parts of the refractive index 
``along the rays". Therefore, non-trivial refractive indices induce 
``distorted" Fourier image of the source.

\section{Demonstration in a simple model}

\begin{figure}
\begin{center}
\includegraphics[width=.9\textwidth]{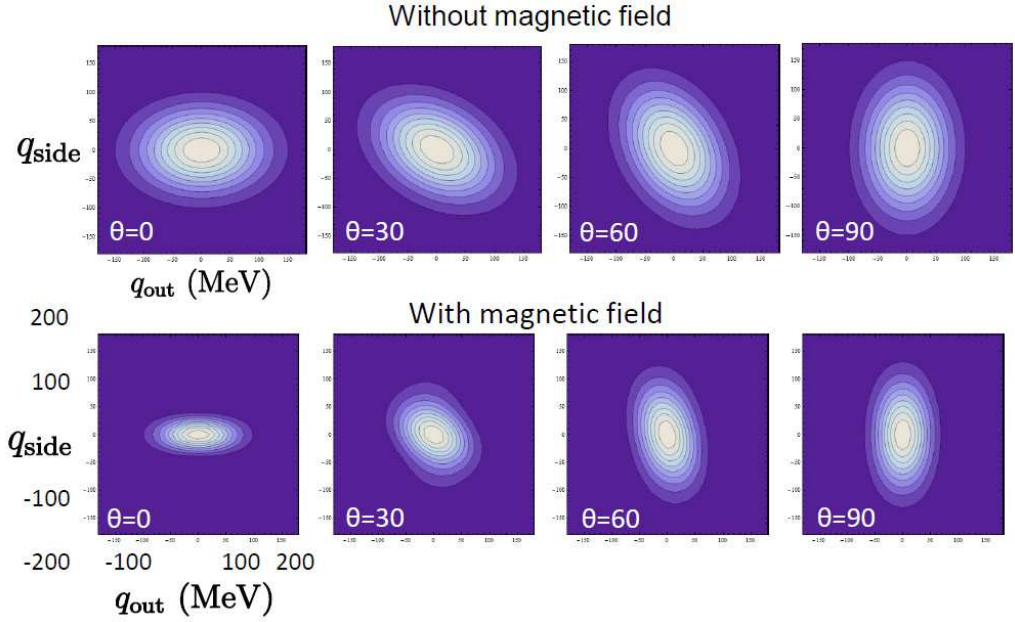}
\end{center}\vspace{-5mm}
\caption{Comparison of photon HBT correlation function. Lower (upper) panels are 
images with (without) the magnetic field. All the panels have the 
same scales.}
\label{fig5}
\end{figure}

Since a complete analytic expression of the refractive indices in a 
magnetic field  is not available so far
(see however Ref.~\cite{HattoriItakura}), 
we consider a simple model which captures qualitative aspects of 
the refractive indices in a magnetic field. Figure 4 shows 
(the real part of) 
the refractive index $n_1$ in the lowest-Landau-level 
approximation for $B/B_c=100$. 
The angle corresponds to the $\theta$ shown in Fig.~2, and 
the distance from the origin corresponds to $|n_1|$. 
Photon energy is taken just below 
the threshold (of the $e^+e^-$ decay) $\omega^2/4m_e^2=0.99$. 
If there is no variation in the refractive index, it should be 
 a unit circle. 
However, as clearly seen in this figure, there is a 
deviation from unity especially in the direction perpendicular to 
the magnetic field.
This can be simply modelled as 
$n=1+\delta n\cdot \exp\{-\frac{x^2}{2\sigma_x^2}-\frac{y^2}{2\sigma_y^2} \}$. 
For simplicity, we use $\sigma_x=10$~fm, $\sigma_y=5$~fm, and 
$\delta n =0.5$ for the model.  
Let us further use the Gaussian source $S(\bx)\propto \exp 
\{ -\frac{x^2}{2R_x^2} -\frac{y^2}{2R_y^2}-\frac{z^2}{2R_z^2}\}$.
Then, in the absence of the magnetic field, the correlation function 
is simply given by its Fourier transform:
$C(\bk, q_{out}, q_{side},0)=1+ \exp \{-R_x^2q_x^2-R_y^2q_y^2\}$ where 
$q_{out}$ and $q_{side}$ are outward and sideward momenta, respectively.
Thus, in the absence of the magnetic field, the HBT correlation function is again 
the Gaussian,
and the shape does not change according to the angle $\theta$. This is 
shown in the upper panels of Fig.~5. On the other hand, the results with 
the magnetic field are
shown in the lower panels of Fig.~5. 
Clearly, one can see distortion of the shape compared to the upper panels, 
and shrinks of the correlation ranges in the momentum space, 
which implies magnifications of the source images
in the coordinate space (similar to lenzing). 
One can also include the imaginary part in the refractive index, which
implies the decay of a photon into an $e^-e^+$ pair. Imaginary part enters 
through the extinction coefficient $\Gamma(\bx;\bk,\bq)$.

\section{Summary}

We have demonstrated in a simple model that the photon HBT image will be 
distorted by effects of the vacuum birefringence which become significant 
in the strong magnetic field. 
If the photon HBT is measured in the future experiments, one has to take 
the effects of distortion into account to obtain correct information of the
matter profile at the early-time stages of the heavy-ion collision. 
Detailed analysis, which goes beyond the simple model as shown in this talk, 
will be reported elsewhere \cite{future}.

\section*{Acknowledgments}

The authors thank the organizers for giving an opportunity to talk 
in the conference.

\end{document}